\documentclass[twocolumn,showpacs]{revtex4}
\usepackage{amssymb}
\usepackage{amsmath}
\usepackage{graphicx}
\usepackage{bm}

\begin{document}

\author{Z. Z. Zhang}
\author{Kai Chang}
\email{kchang@red.semi.ac.cn}
\affiliation{SKLSM, Institute of
Semiconductors, Chinese Academy of Sciences, P.O. Box 912, Beijing
100083, China}
\author{K. S. Chan}
\affiliation{Department of Physics
and Materials Science, City University of Hong Kong, Hong Kong,
China}
\date{\today }
\title{Resonant Tunneling through double-bended Graphene Nanoribbons }

\begin{abstract}
We investigate theoretically resonant tunneling through double-bended
graphene nanoribbon structures, i.e., armchair-edged graphene nanoribbons
(AGNRs) in between two semi-infinite zigzag graphene nanoribbon (ZGNR)
leads. Our numerical results demonstrate that the resonant tunneling can be
tuned dramatically by the Fermi energy and the length and/or widths of the
AGNR for both the metallic and semiconductor-like AGNRs. The structure can
also be use to control the valley polarization of the tunneling currents and
could be useful for potential application in valleytronics devices.
\end{abstract}

\maketitle

Graphene is a single layer of carbon atoms arranged in a hexagonal lattice.
Recently, graphene samples have been fabricated experimentally by
micro-mechanical cleavage of graphite \cite{Novoselov}. This material has
aroused the increasing attention due to its novel transport property that
arises from its unique band structure: the conduction and valence bands meet
conically at the two nonequivalent Dirac points, called $K$ and $K^{^{\prime
}}$ valleys, of the Brillouin zone, which show opposite chirality. Around
the two points (called Diract points), the energy dispersion is linear and
described by the massless Dirac equation. In graphene, the presence of edges
can change the energy spectrum of the $\pi $-electron dramatically. Graphene
nanoribbons (GNRs) have been fabricated by using conventional lithography
and etching techniques \cite{IBM,kim}. The electronic properties of a GNR
depend very sensitively on the size and shape of edges, i.e., zigzag- and
armchair-edged GNR. The zigzag-edged graphene nanoribbons (ZGNRs) and
armchair-edged graphene nanoribbons (AGNRs) exhibit different band
structures. For the ZGNRs, there are always the localized states appearing
at the edge near the Dirac point. Therefore the ZGNRs exhibit metallic-like
behavior. The AGNRs show metallic-like and semiconductor-like features
alternatively as the width of nanoribbons increases \cite{Nakada}. Those
features are very different from the conventional semiconductor quantum
wire, provide possible ways to tailor the transport and optical properties
of GNR \cite{Wakabayashi}, and pave a new path to potential applications of
valleytronics device, e.g., the quantum point contact is used to realize a
valley filter and a valley valve utilizing the edge state of ZGNRs \cite%
{Beenakker}. Interestingly, the valleys $K$ and $%
K^{^{\prime }}$ are decoupled for ZGNRs, but mixed for AGNRs \cite{Brey}. It
is natural to ask and image what happens when we construct a mesoscopic
device by combining the ZGNRs\ and AGNRs.

\begin{figure}[tbp]
\includegraphics [width=\columnwidth]{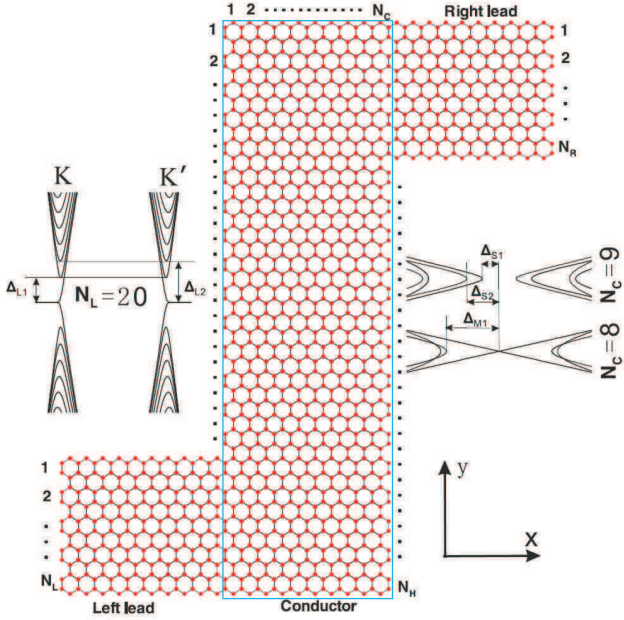}
\caption{(Color online) Schematic diagram of the proposed graphene
nanoribbon structure. The left (right) inset describes the energy band
structure for the left and right lead (AGR). The blue frame denotes the
region of middle conductor, which consists of the AGNR with the length $%
N_{H} $ and width $N_{c}$.}
\label{fig1}
\end{figure}

In this work, we investigate theoretically the resonant tunneling through
double-bended GNR structures, i.e., a AGNR in between two ZGNR leads (see
Fig. \ref{fig1}). For transport through metallic-like AGNRs, electrons can
not transmit perfectly but show a resonant tunneling behavior. For
semiconductor-like AGNRs, the resonant tunneling is blocked when the Fermi
energy $E_{F}$ is lower than the bandgap of AGNRs and displays similar
resonant peaks when the Fermi energy $E_{F}$ exceeds the gap. Our
theoretical results show that this kind of structure can control the valley
polarization of tunneling currents.

The electronic states in graphene are described by a nearest-neighbor
tight-binding Hamiltonian $H=\sum\limits_{<i,j>}t_{ij}c_{i}^{\dagger }c_{j}$%
, where $t_{ij}=-t$ ($t=3.03$ eV) is the transfer energy of the
nearest-neighbour sites\cite{Saito}. The conductance of the system is
evaluated using the multichannel Landauer formula\cite{butiker},
\begin{equation}
G\left( E_{F}\right) =\frac{e^{2}}{\pi \hslash }\sum\limits_{\mu }T_{\mu
}\left( E_{F}\right) ,T_{\mu }\left( E_{F}\right) =\sum\limits_{\nu
}\left\vert t_{\mu ,\nu }\left( E_{F}\right) \right\vert ^{2},  \label{eq1}
\end{equation}%
where $t_{\mu ,\nu }\left( E_{F}\right) $ is a transmission coefficient from
$\nu $-th channel in the left lead to $\mu $-th channel in the right lead at
the Fermi energy $E_{F}$, calculated by a recursive Green's function method%
\cite{Ando}. As schematically shown in Fig. \ref{fig1}, an AGNR is connected
to two metallic semi-infinite ZGNR leads. The insets in Fig. \ref{fig1} show
the energy band structures of the left (right) ZGNR lead and middle AGNR.
The Fermi energy $E_{F}$ can be tuned experimentally through the electric
top- or back-gates\cite{Ozyilmaz}. In our calculation, all physical
quantities are introduced dimensionlessly, e. g., the energy $E$ and the
conductances are in units of $t$ and $e^{2}/\left( \pi \hslash \right) $,
respectively.

\begin{figure}[tbp]
\includegraphics [width=\columnwidth]{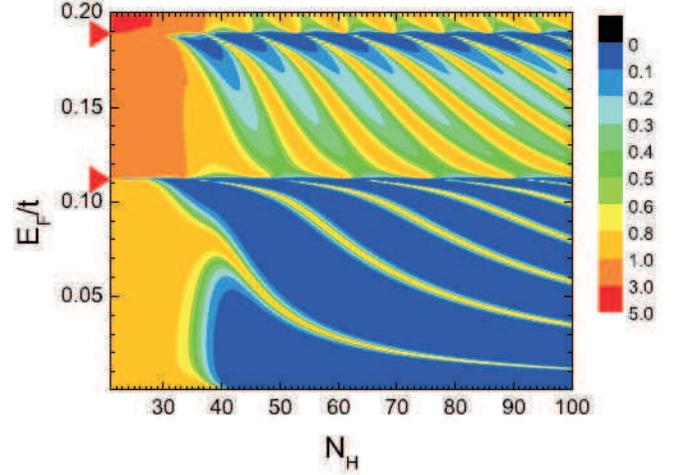}
\caption{(Color online) The contour plot of the conductance as a function of
the length of the AGR ($N_{H}$) and Fermi energy ($E_{F}$). $N_{L}=N_{R}=20$%
, $N_{A}=8$. The red triangles indicate the energies which corresponding the
bottoms of the second and third transverse subbands in the left or right
lead. }
\label{fig2}
\end{figure}

First, we consider that the middle conductor of the double-bended GNRs
consists of a metallic-like AGNR with $N_{C}=8$ and $\Delta _{M1}=0.286$
(see the right inset of Fig.1). The left and right leads have the same width
$N_{L}=N_{R}=20$ with $\Delta _{L1}=0.113$ and $\Delta _{L2}=0.189$ (see the
left inset of Fig.1). Fig. \ref{fig2} shows the contour plot of the
conductance as functions of the length of the middle AGNR ($N_{H}$) and the
Fermi energy ($E_{F}$). Fig. \ref{fig2} demonstrates that there are three
regimes indicated by the red triangles. The three regimes correspond to the
opening of the second and third transverse modes. When $N_{H}=N_{L}=20$, the
system will return to a perfect ZGNR and the conductance exhibits a
step-like feature, i.e., $1,3,5$ $\cdots (e^{2}/\pi \hslash )$. When $%
N_{H}>N_{L}+N_{R}$, the resonant tunneling peaks appear regularly as the
Fermi energy increases and become more and more as the length of AGNR region
$N_{H}$ increases. This resonant tunneling behavior arises from the
constructive interference effect when the electron wave propagates back and
forth in the AGNR region. The incident electron can be completely reflected
due to the destructive interference, especially when the energy of the
incident electron is lower than the onset of the lowest mode of the left
(right) lead, i.e., $E<\Delta _{L1}$, though the middle AGNR is metallic
with zero energy gap. For the fixed length, the resonance peaks broaden as
the Fermi energy increase due to the enhanced coupling between the electron
states in the AGNR and that in the ZGNR leads. When $\Delta _{L1}<E<\Delta
_{L2},$ there are three propagating modes, two of which belong to the $K$
valley and one mode belongs to the $K^{^{\prime }}$ valley. The similar
oscillation can also be see when the incident energy is higher than the
onset of the second modes in the left (right) lead.

\begin{figure}[tbp]
\includegraphics [width=\columnwidth]{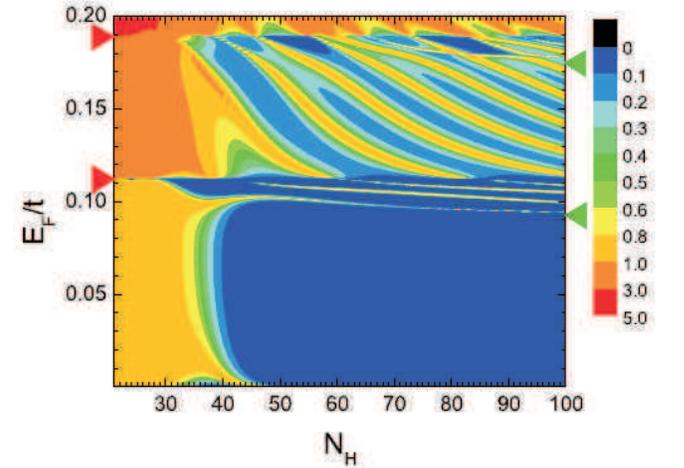}
\caption{(Color online) The same as Fig. \protect\ref{fig2}, but for $%
N_{C}=9 $ AGR conduct. The green triangles indicate the energies
corresponding to $\Delta _{S1} $ and $\Delta _{S2}$} \label{fig3}
\end{figure}

Next, we consider the middle AGNR ($N_{C}=9$, $\Delta _{S1}=$ $0.092,$ and $%
\Delta _{S2}=0.176,$ see the right inset of Fig. 1) showing the
semiconductor-like feature. When $N_{H}$ approaches to $N_{L}$, the
situation is the same as for $N_{C}=8$ and show perfect transmission. But
when $N_{H}$ increases and is larger than $N_{L}+N_{R}$, electron tunneling
can be fully blocked for $0<E<\Delta _{S1}$. This phenomena reveals that
semiconductor AGNR behaves like an opaque barrier and can confine the
electron between two AGNRs. For $\Delta _{S1}<E$ $<\Delta _{L1}$ (see the
left inset of Fig.1), similar resonance tunneling peaks appear, but are
sharper than those through the metallic-like AGNR. When the incident energy
increases further and is higher than $\Delta _{S2}$, the resonant tunneling
becomes more complicated, and even shows crossing and anti-crossing features
that are not found for the metallic AGNRs (see Fig. 2). The crossing and
anticrossing behaviors are caused by the existence of the higher transverse
modes of electron in the semiconductor-like AGNR region when the Fermi
energy $E_{F}>\Delta _{S2}$ (see the right inset of Fig.1). The absence of
the crossing and anticrossing behavior for the metallic AGNR (see Fig. 2)
arises from that the metallic AGNR only supports one channel when $%
E_{F}<\Delta _{M1}$ (see the right inset of Fig.1).
\begin{figure}[tbp]
\includegraphics [width=\columnwidth]{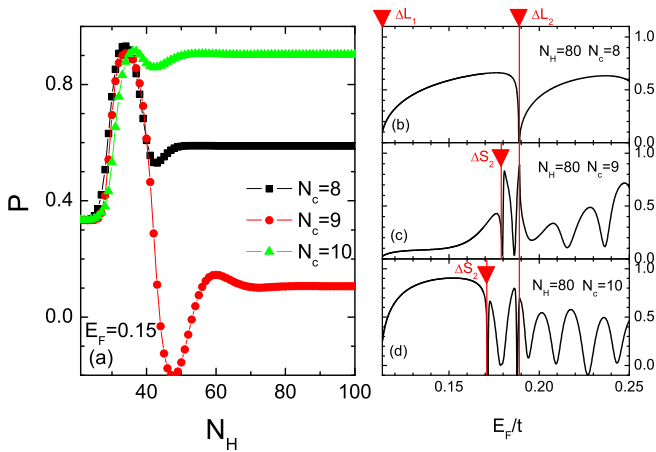}
\caption{(Color online) (a) Valley polarization (P) of the tunneling current
as a function of the length ($N_{H}$) of the middle conductor for the three
different widths at the fixed chemical potential $E_{F}=0.15$. (b), (c), and
(d) Valley polarization as a function of the chemical potential ($E$) at the
fixed length ($N_{H}=80$) of middle AGNRs for $N_{C}=8$, 9, and 10,
respectively. }
\label{fig4}
\end{figure}

Fig. \ref{fig4}(a) shows the valley polarization of the tunneling current as
a function of the length of AGNR for the different widths ($N_{C}$) of the
AGNRs, where the valley polarization $P$ is defined as $P=(\sum\limits_{\mu
\in K}T_{\mu }-\sum\limits_{\mu \in K^{\prime }}T_{\mu })/\sum\limits_{\mu
}T_{\mu }$ \cite{Beenakker}. There are the three channels propagating modes (%
$N_{m}=3$) in the leads along the \textit{x} axis at $E_{F}=0.15$. Two
channels belong to the $K$ valley and the other belongs to the $K^{^{\prime
}}$ valley. When the length of the AGNR approaches to the lead width, i.e., $%
N_{H}=N_{L}$, the structures become a perfect ZGNR and the valley
polarization is equal to $1/N_{m}$($N_{m}=3$) since there is no coupling
between the $K$ and $K^{^{\prime }}$ valleys. When $N_{H}>N_{L}$(or $N_{R})$%
, the transmitted electron no longer belongs to the $K$ valley purely when
the incident electron belonging to $K$ goes through AGNR due to the coupling
between $K$ and $K^{^{\prime }}$ valleys in the AGNR. As the length of the
AGNR increases, the polarization increases very fast and reaches the maximum
at $N_{H}\approx 35$. When the length increases further, the valley
polarizations decrease and saturate at specific values. The AGNRs with
different widths have different saturated values. The AGNR with $N_{c}=3n+1$
($n=3$) show the highest saturated valley polarization, approaching to $0.9$%
. For the AGNRs with $N_{c}=3n-1$, the saturated value is the lower. The
saturated polarization of the AGNR with $N_{c}=3n$ is the lowest and
approaches to $0.1$. Figs. \ref{fig4}(b), (c), and (d) show the valley
polarizations as a function of the Fermi energy for different widths of
AGNRs. When there is a single channel in the middle AGNR region, i.e., $%
E_{F}<\Delta _{M1}$, for the metallic-like AGNR$\,$ and $E_{F}<\Delta _{S2}$
for the semiconductor-like AGNR, the valley polarization varies smoothly.
Once the Fermi energy exceeds the critical energies ($\Delta _{M1}$ or $%
\Delta _{S2}$), the valley polarization oscillates heavily (see Figs. \ref%
{fig4}(c) and (d)) because of the interference between different channels in
the AGNR region and ZGNR leads which are opened orderly with increasing the
Fermi energy. But the valley polarization in Fig. 4(b) still changes
smoothly as the Fermi energy increases. This is because the AGNR only
supports a single channel when the energy of the second channel is higher
than the Fermi energy, i.e., $\Delta _{M1}>E_{F}$ (see the right inset of
Fig. 1). For the single incident channel case, i.e., $0<E_{F}<\Delta _{L1}$,
the valley polarization of tunneling current is always equal to $1$ since
there is a single channel belonging to the $K$ valley. From Fig. \ref{fig4},
one can see that the valley polarization of the tunneling current can be
changed dramatically by tuning the Fermi energy and the length of the AGNR.

In summary, we have investigated theoretically resonant tunneling through a
double-bended GNR, i.e., an AGNR in between two ZGNR leads. Our numerical
results demonstrate that the resonant tunneling can be tuned dramatically by
the Fermi energy and the length and/or widths of the AGNR for both the
metallic and semiconductor-like AGNRs. The valley polarization saturate as
the length of the AGNRs increases, and the saturated valley polarizations
depend sensitively on the widths of the AGNRs. The structure we proposed can
be used to manipulate the valley polarization of the tunneling current and
should be useful for potential application in valleytronics devices.

\begin{acknowledgments}
This work is supported by the NSF of China Grant No. 60525405.
\end{acknowledgments}

\end{document}